\pdfoutput=1  

\documentclass[aps,pre,10pt,superscriptaddress,floatfix,notitlepage]{revtex4-1}

\usepackage[english]{babel}
\usepackage{graphicx}
\usepackage{amsmath, amssymb} 
\usepackage{textcomp}
\usepackage{gensymb}  
\usepackage{bm}  

\newcommand{\Ra}{\textit{Ra}}  
\renewcommand{\Re}{\textit{Re}}  
\renewcommand{\Pr}{\textit{Pr}}  
\newcommand{\Pm}{\textit{Pm}}  
\newcommand{\Nu}{\textit{Nu}}  
\newcommand{\Ha}{\textit{Ha}}  
\newcommand{\Rm}{\textit{Rm}}  
\newcommand{\N}{\textit{N}}  

\begin{document}

\title{Flow regimes of Rayleigh-B\'enard convection in a vertical magnetic field}

\author{Till Z\"urner}
\email{till.zuerner@tu-ilmenau.de}
\affiliation{Institute of Thermodynamics and Fluid Mechanics, Technische Universit\"at Ilmenau, Postfach~100565, D-98684 Ilmenau, Germany}

\author{Felix Schindler}
\affiliation{Department of Magnetohydrodynamics, Institute of Fluid Dynamics, Helmholtz-Zentrum Dresden - Rossendorf, Bautzner Landstra\ss e 400, D-01328 Dresden, Germany}

\author{Tobias Vogt}
\affiliation{Department of Magnetohydrodynamics, Institute of Fluid Dynamics, Helmholtz-Zentrum Dresden - Rossendorf, Bautzner Landstra\ss e 400, D-01328 Dresden, Germany}

\author{Sven Eckert}
\affiliation{Department of Magnetohydrodynamics, Institute of Fluid Dynamics, Helmholtz-Zentrum Dresden - Rossendorf, Bautzner Landstra\ss e 400, D-01328 Dresden, Germany}

\author{J\"org Schumacher}
\affiliation{Institute of Thermodynamics and Fluid Mechanics, Technische Universit\"at Ilmenau, Postfach~100565, D-98684 Ilmenau, Germany}

\begin{abstract}
The effects of a vertical static magnetic field on the flow structure and global transport properties of momentum and heat in liquid metal Rayleigh-B\'enard convection are investigated.
Experiments are conducted in a cylindrical convection cell of unity aspect ratio, filled with the alloy GaInSn at a low Prandtl number of $\Pr=0.029$.
Changes of the large-scale velocity structure with increasing magnetic field strength are probed systematically using multiple ultrasound Doppler velocimetry sensors and thermocouples for a parameter range that is spanned by Rayleigh numbers of $10^6 \le \Ra \le 6\times 10^7$ and Hartmann numbers of $\Ha \le 1000$.
Our simultaneous multi-probe temperature and velocity measurements demonstrate how the large-scale circulation is affected by an increasing magnetic field strength (or Hartmann number).
Lorentz forces induced in the liquid metal first suppress the oscillations of the large-scale circulation at low~$\Ha$, then transform the one-roll structure into a cellular large-scale pattern consisting of multiple up- and downwellings for intermediate~$\Ha$, before finally expelling any fluid motion out of the bulk at the highest accessible~$\Ha$ leaving only a near-wall convective flow that persists even below Chandrasekhar's linear instability threshold.
Our study thus proves experimentally the existence of wall modes in confined magnetoconvection.
The magnitude of the transferred heat remains nearly unaffected by the steady decrease of the fluid momentum over a large range of Hartmann numbers.
We extend the experimental global transport analysis to momentum transfer and include the dependence of the Reynolds number on the Hartmann number.
\end{abstract}

\maketitle

\section{Introduction}
\label{sec:intro}

The coupling of magnetic fields to thermal convection in electrically conducting fluids can result in profound changes of the flow patterns and transport properties.
In nature, this fluid motion, which is termed magnetoconvection, prominently occurs in the interiors of stars or the liquid metal cores of planets, where convective flows and global magnetic fields create the complex dynamo effect~\citep{Davidson2001,Moffatt2019}.
Technological applications include the flow control of hot metal melts by electromagnetic brakes in metallurgy or the influence of strong magnetic fields on liquid metal cooling systems proposed for fusion reactor blankets~\cite{Ihli2008}.
A better understanding of these systems can be achieved by experimental studies in canonical configurations.
Here, we investigate the case of Rayleigh-B\'enard convection (RBC) with an imposed vertical static magnetic field~$B_0$, one of the elementary magnetoconvection configurations.
RBC considers a horizontal fluid layer of height~$H$, heated from below and cooled from above.
It is governed by three dimensionless numbers which are the Rayleigh number~$\Ra$, the Prandtl number~$\Pr$ and the aspect ratio~$\Gamma$.
Magnetoconvection adds two more dimensionless parameters -- the Hartmann number $\Ha$ and the magnetic Prandtl number $\Pm$.
The aspect ratio~$\Gamma$ is the horizontal extent of the fluid layer normalized by its height~$H$.
The remaining four parameter are given by
\begin{equation}
\Ha = B_0 H \sqrt{\frac{\sigma}{\rho\nu}} \,, \quad
\Pm = \mu_0\sigma\nu \,, \quad
\Pr = \frac{\nu}{\kappa} \,, \quad
\Ra = \frac{g \alpha \Delta T H^3}{\nu \kappa} \,,
\end{equation}
with the electrical conductivity~$\sigma$, the permeability of free space~$\mu_0$, the mass density~$\rho$, the kinematic viscosity~$\nu$, the temperature diffusivity~$\kappa$, the acceleration due to gravity~$g$, the thermal expansion coefficient~$\alpha$, and the temperature difference~$\Delta T$ between bottom and top boundary of the fluid layer.

The effect of the external magnetic field on the convective flow depends on its strength, quantified by the Hartmann number, and its orientation towards gravity.
Canonical configurations apply a static homogeneous magnetic field either in horizontal or in vertical direction.
For a horizontal field -- perpendicular to gravity -- the induced Lorentz forces organise the convective flow in quasi-two-dimensional rolls which are aligned in the direction of the field~\cite{Yanagisawa2013}.
The number of rolls depends on the ratio $\Ra/\Ha^2$.
The flow experiences a multitude of dynamical phenomena during the transition between different roll configurations~\cite{Yanagisawa2011,Tasaka2016} or when transitioning towards fully three-dimensional turbulence for lower $\Ha$~\cite{Vogt2018}.
In contrast, vertical magnetic fields -- parallel to gravity -- are known to suppress the convective flow as shown in theory~\cite{Chandrasekhar1961,Houchens2002,Busse2008}, direct numerical simulations (DNS) at high and low Prandtl numbers~\cite{Liu2018,Lim2019,Yan2019} and experiments~\cite{Nakagawa1955,Cioni2000,Burr2001,Aurnou2001,King2015}.
For this case, the linear instability threshold of an infinitely extended magnetoconvection layer itself depends on the Hartmann number and has been determined by \citet{Chandrasekhar1961}.
At high $\Ha$ it can be estimated by $\Ra_\mathrm{c} \approx \pi^2 \Ha^2$ or in turn, the magnetic field strength below which convective fluid motion is observed follows to $\Ha_\mathrm{c} \approx \sqrt{Ra}/\pi$.
This approximation is often referred to as the Chandrasekhar limit (which should not be confused with the one for the maximum mass of a stable white dwarf star).
\citet{Chandrasekhar1961} determined $\Ha_\mathrm{c}$ by a linear stability analysis in an horizontally unbounded fluid layer.
As such, $\Ha_\mathrm{c}$ does not include the effect of side-walls, which will be discussed in section~\ref{sec:regimemap}.

Magnetoconvection experiments are generally conducted using liquid metals, which have a very low Prandtl number $\Pr \ll 1$ and a high electrical conductivity $\sigma \sim 10^6$\,S\,m$^{-1}$.
Recently, magnetoconvection experiments at a higher Prandtl number $\Pr = 12$ were conducted using sulphuric acid~\cite{Aujogue2016}.
While this setup allows the use of optical flow measurement methods, very high magnetic fields of $B_0\sim 10$ T are necessary to reach the same parameters as in liquid metals.

Here, we report experimental investigations of liquid metal convection in a closed cylindrical container of unit aspect ratio in the presence of a uniform vertical magnetic field~$B_0$, an extension of our previous studies which were reported in \citet{Zurner2019}.
The application of multiple thermocouple probes for temperature measurement as well as multiple crossing ultrasound beam lines allows us to reconstruct various regimes of the large-scale circulation flow in dependence on the external magnetic field strength (i.e., the Hartmann number $\Ha$) and the strength of thermal driving of the flow (i.e., the Rayleigh number $\Ra$).
Our work extends previous experimental studies (which are listed above) in several points. 
(1)~We provide a systematic survey of the different flow regimes in the $(\Ra, \Ha)$ parameter plane which is impossible in DNS at low Prandtl numbers for longer time spans.
(2)~We give an experimental evidence for the existence of wall modes in magnetoconvection, remnants of the convection rolls or cells that maintain a convective heat transfer across the vessel for Rayleigh numbers below the linear instability threshold $\Ha_\mathrm{c}$ and thus confirm recent results of \citet{Liu2018}.
The appearance and structure of these modes has been predicted theoretically by \citet{Houchens2002} and \citet{Busse2008}.
(3)~The comprehensive usage of ultrasound velocimetry makes a direct determination of the momentum transfer in the flow possible for the first time which extends the analysis in previous experiments.
Both, global heat and momentum transfer as a function of the Hartmann number can thus be monitored in detail and compared with existing data of experiments and DNS.
   
This article is structured as follows.
The next section presents the experimental set-up and measurement techniques.
Section~\ref{sec:large-scale} shows the changes in the flow structure with progressively larger magnetic field strength.
The global transport properties of heat and momentum in the convective system are discussed in section~\ref{sec:transport} in light of the determined flow regimes.
Finally, we summarize our findings in section~\ref{sec:conclusion} and give an outlook.
The measurement results of selected quantities from the present article are also enclosed as supplementary material.

\section{Experimental set-up}
\label{sec:setup}
\begin{figure}
\centering
\includegraphics{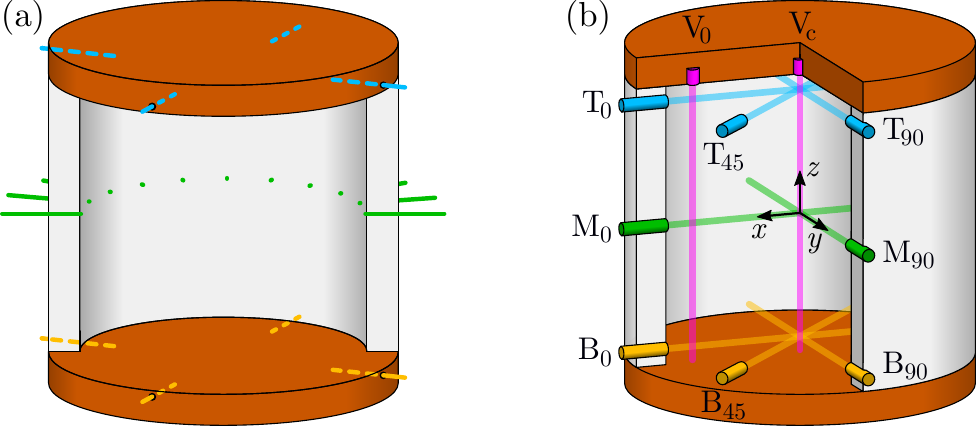}
\caption{%
  Positions of the measurement devices.
  (a)~Thermocouples for the temperature measurement.
  Four sensors are mounted in the top plate (blue), four in the bottom plate (orange) and eleven sensors in form of an array at mid height (green).
  (b)~UDV sensors for the velocity measurement. 
  Radial velocity components are taken near the top plate (label~T, blue) at mid height (label~M, green) and near the bottom plate (label~B, orange).
  The vertical velocity component (purple) is taken along the cylinder axis (label~V\textsubscript{c}) and at a radial position of $r/R = 0.8$ (label~V\textsubscript{0}).
  The label subscripts refer to the azimuthal position of the sensors in degree.}
\label{fig:setup}
\end{figure}
The set-up is the same as in the experiments by \citet{Zurner2019}, extended by the magnetic field generation components. For completeness, we provide a compact description in the following.
Measurements are conducted in a cylindrical cell of height~$H=180\,$mm and diameter~$D=180\,\text{mm} = 2R$, giving an aspect ratio~$\Gamma = D/H = 1$ as shown in  figure~\ref{fig:setup}.
The top and bottom plate are made of copper.
Heat is supplied from below by an electrical heating pad.
The top is cooled using water from a thermostat.
The working fluid is the liquid metal alloy GaInSn with a melting temperature of 10.5\,\degree C.
At 35\,\degree C it has an electrical conductivity of $\sigma = 3.2 \times 10^6\,\text{S}\,\text{m}^{-1}$~\cite{Plevachuk2014}, allowing for a large effect of magnetic fields on the flow.
The relevant diffusivities of the alloy are the kinematic viscosity~$\nu = 3.2\times10^{-7}\,\text{m}^2\,\text{s}^{-1}$, the thermal diffusivity~$\kappa = 1.1\times10^{-5}\,\text{m}^2\,\text{s}^{-1}$ and the magnetic diffusivity~$\eta = (\mu_0\sigma)^{-1} = 0.25\,\text{m}^2\,\text{s}^{-1}$.
These values yield a Prandtl number $\Pr = 0.029$ and a magnetic Prandtl number $\Pm = 1.36 \times 10^{-6}$.
The vertical magnetic field is generated by the MULTIMAG facility at the Helmholtz-Zentrum Dresden-Rossendorf~\cite{Pal2009} with a maximum magnetic flux density of $B_0 = 0.14$\,T.
With this set-up, a Rayleigh number range of $10^6 \le \Ra \le 6\times10^7$ and a Hartmann number range of $0 \le \Ha \le 1000$ is achieved.

The temperatures $T_\mathrm{top}$ of the top plate and $T_\mathrm{bot}$ of the bottom plate are measured using four thermocouples each, distributed evenly around the circumference and 4\,mm from the liquid (figure~\ref{fig:setup}(a)).
These probes are used to determine the temperature difference between the plates $\Delta T = T_\mathrm{bot} - T_\mathrm{top}$ and the average fluid temperature $\bar T = (T_\mathrm{top} + T_\mathrm{bot})/2$.
The temperature of the fluid at the side-wall is measured by a thermocouple array of eleven sensors at mid-height of the cell.
They cover half the circumference and are used to detect hot up- and cold downwelling fluid.
Finally, the temperature of the cooling water is measured at the in- and the out-flow of the upper copper plate ($T_\mathrm{in}$ and $T_\mathrm{out}$).
Together with the volume flux~$\dot V$ of the water, the heat flux through the fluid is $\dot Q = \tilde c_p \tilde\rho \dot V (T_\mathrm{out} - T_\mathrm{in})$.
Here, $\tilde c_p$ is the isobaric heat capacity and $\tilde\rho$ the mass density of water~\cite{Cengel2008}.
Only measurements with $T_\mathrm{out} - T_\mathrm{in} > 0.2$\,K are considered.
The heat flux is additionally corrected by the heat loss through the side-walls using three pairs of thermocouples which measure the radial temperature gradient in the side-wall.
More details can be found in \citet{Zurner2019}.

The flow field in the container is probed using ultrasound Doppler velocimetry (UDV).
Ten sensors are distributed around the cell as depicted in figure~\ref{fig:setup}(b).
The radial velocity component is recorded at three vertical positions: 10\,mm below the top plate, at mid-height and 10\,mm above the bottom plate.
Near the plates, three sensors are placed at azimuthal positions $\phi = 0\degree$, $45\degree$ and $90\degree$ (sensors T\textsubscript{0}, T\textsubscript{45} and T\textsubscript{90} at the top and B\textsubscript{0}, B\textsubscript{45} and B\textsubscript{90} at the bottom).
At mid height sensors M\textsubscript{0} and M\textsubscript{90} are positioned at $\phi = 0\degree$ and $90\degree$, respectively.
Additionally, the vertical velocity component is measured by two sensors along the central axis of the cell~(V\textsubscript{c}) and at radial position $r/R = 0.8$ and azimuthal position $\phi = 0\degree$~(V\textsubscript{0}).

The experiments are started by setting the temperature difference between the plates to $\Delta T = 0$\,K until no movement of the fluid can be detected.
Next, the desired temperature difference is approached and held, until the flow has settled again.
Finally, the magnetic field is applied.
Once the top and bottom temperature and the flow have reached a steady state, the measurement is started.
When changing to a new magnetic field strength, the experiment is reset to $B=0$\,T first.
The average fluid temperature is kept constant at $\bar T = 35$\,\degree C except at the highest $\Ra = 6\times10^7$, where it increases to $\bar T \sim 40$\,\degree C for $Ha < 20$, resulting in $\Pr=0.028$ and $\Pm = 1.34 \times 10^{-6}$.

\section{Hartmann number dependence of the large-scale flow}
\label{sec:large-scale}

Without external magnetic field, the turbulent flow in our cell organizes into a single convection roll, the so-called large-scale circulation (LSC)~\cite{Ahlers2009,Chilla2012,Zurner2019}.
This structure has been extensively investigated in water experiments~\citep{Funfschilling2004,Sun2005,Brown2006,Xi2009,Zhou2009,Xie2013}.
The opaque nature of liquid metals makes it much more difficult to study their internal flow structure.
Earlier experiments thus mainly focussed on temperature characteristics and velocity measurements using single-UDV probes or local temperature correlations~\cite{Takeshita1996,Cioni1997,Glazier1999,Tsuji2005,King2013}.
Only recently the has the global flow structure of liquid metal convection been studied in more detail~\cite{Khalilov2018,Vogt2018a,Zurner2019}.
The LSC in turbulent convection at $\Gamma=1$ exhibits a variety of dynamical properties.
The torsion and sloshing modes refer to short-term regular oscillations ($\sim 10 \tau_\mathrm{ff}$ with the free-fall convective time unit $\tau_\mathrm{ff}=\sqrt{H/(g\alpha\Delta T)}$), which deform the shape of the convection roll into a three-dimensional flow.
Over longer time-scales ($\sim 100$ to $1000 \tau_\mathrm{ff}$) the average azimuthal orientation of the LSC drifts due to the axis-symmetry of the system.
This meandering motion follows the physical laws of a diffusion process as shown by \citet{Brown2006}.
Reversals and cessations of the LSC appear on even larger time-scales~\cite{Brown2006}.
For a more detailed analysis of the classical RBC system in the present set-up we refer to our previous work~\cite{Zurner2019}.
The following subsections describe how the large-scale flow structure of the convection system is altered by the external vertical magnetic field $B_0$ with a successively stronger magnitude.
The boundaries between the different resulting flow regimes are summarized in section~\ref{sec:regimemap}.

\subsection{The large-scale one-roll regime}
\label{sec:LSC}
\begin{figure}
\centering
\includegraphics{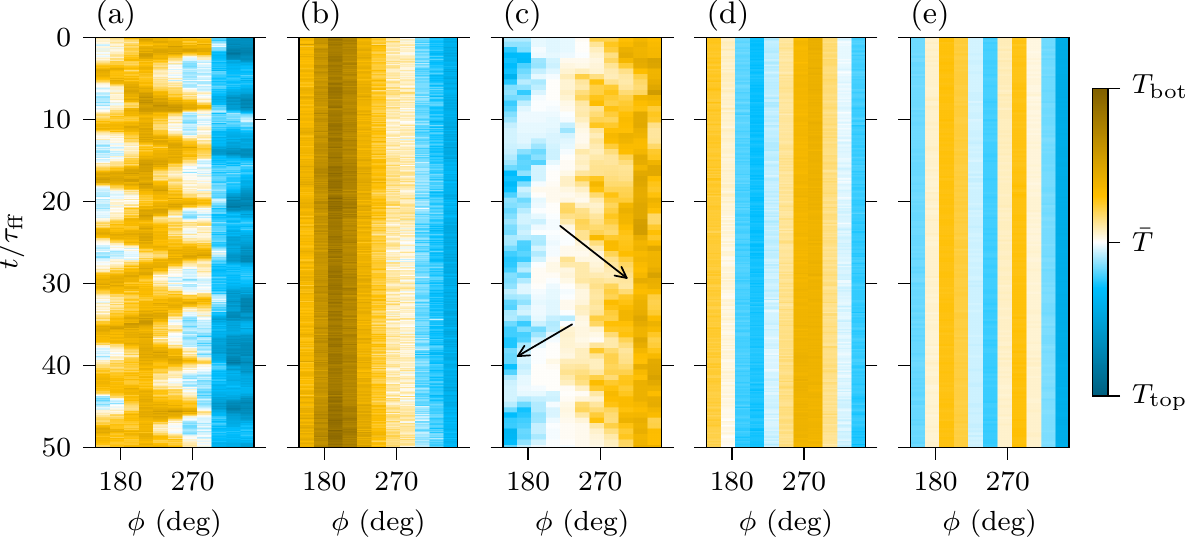}
\caption{%
  Side-wall temperature profiles for different vertical magnetic field strengths.
  The space-time plots at half-height of the RBC flow are reconstructed from the thermocouple array.
  The temperature is colour-coded and plotted over the azimuthal position~$\phi$ and time~$t$ in free-fall time units~$\tau_\mathrm{ff}$.
  (a)~LSC with regular oscillations ($\Ra = 10^6$, $\Ha = 13$, $\tau_\mathrm{ff} = 17.2$\,s).
  (b)~LSC without oscillations ($\Ra = 10^6$, $\Ha = 131$, $\tau_\mathrm{ff}=17.2$\,s).
  (c)~LSC with plume entrainment ($\Ra = 6\times 10^7$, $\Ha = 263$, $\tau_\mathrm{ff} = 2.3$\,s).
  The arrows in this panel indicate the drift of secondary plume positions towards the main convection roll.
  (d)~Cellular pattern ($\Ra = 4.2\times 10^6$, $\Ha = 592$, $\tau_\mathrm{ff} = 8.6$\,s).
  (e)~Wall modes ($\Ra = 1\times 10^7$, $\Ha = 1052$, $\tau_\mathrm{ff} = 5.4$\,s).}
\label{fig:Tmid_Ha}
\end{figure}

For small magnetic field amplitudes, the one-roll structure of the LSC remains intact.
Figure~\ref{fig:Tmid_Ha} displays colour plots of the mid-height temperature over time.
These space-time plots are reconstructed from the signals of the thermocouple array.
At first, the regular oscillations of the sloshing mode are still visible, e.g.,\ for $\Ha=13$ and $\Ra=10^6$ in figure~\ref{fig:Tmid_Ha}(a).
An increase of the magnetic field strength to $\Ha=131$ in figure~\ref{fig:Tmid_Ha}(b) suppresses the oscillation modes and results in an establishment of a  quasi two-dimensional roll.
UDV measurements show that the flow speed successively decreases for higher field strengths which is known from theory~\cite{Chandrasekhar1961,Chakraborty2008,Zurner2016c} and previous experiments~\cite{Cioni1997,King2015}.
This aspect is discussed in detail in section~\ref{sec:transport}.

For higher Rayleigh numbers $\Ra \gtrsim 3\times 10^6$ the flow exhibits a new large-scale roll pattern.
Figure~\ref{fig:Tmid_Ha}(c) at $Ra=6\times10^7$ and $\Ha=263$ shows a dominant LSC roll with a hot up- and cold downwelling flow at $\phi \sim 315\degree$ and $135\degree$, respectively.
Additionally, hot or cold spots appear at the side wall at $\phi \sim 225\degree$.
These indicate localized regions of secondary up- and down-flows that are detached from the primary LSC roll.
They likely consist of plumes which are not transported by the horizontal motion of the LSC at the plates.
However, these secondary flows experience an azimuthal drift and after some time the plumes are again entrained back into the main path of the LSC.
This plume entrainment is highlighted in figure~\ref{fig:Tmid_Ha}(c) for two cases by arrows.
These multiple regions of vertical flow are interpreted as a first sign of the breakdown of the coherent LSC roll structure which cannot be maintained anymore.

In summary, the effect of small external magnetic fields is thus the suppression of the regular oscillation modes of the LSC.
In the following, we will refer to this regime as the LSC regime.

\subsection{The cellular flow regime}
\label{sec:cellular}
As already indicated in the previous section, if the magnetic field magnitude $B_0$ is strong enough, the one-roll LSC structure breaks down into multiple convection rolls or cells.
This can be recognized in the data of the mid-height temperature array as multiple separate hotter and colder regions which correspond to up- and down-flows, respectively, along the side-wall.
For example, figure~\ref{fig:Tmid_Ha}(d) shows two up- and two down-flow regions over half the circumference for $\Ha = 592$ and $\Ra = 4.2\times 10^6$.
A~continuation of this pattern would result in three up- and down-flow regions along the whole circumference,~i.e., a three-fold azimuthal symmetry.

\begin{figure}
\centering
\includegraphics[trim={0.95cm 0 0 0}, clip=true]{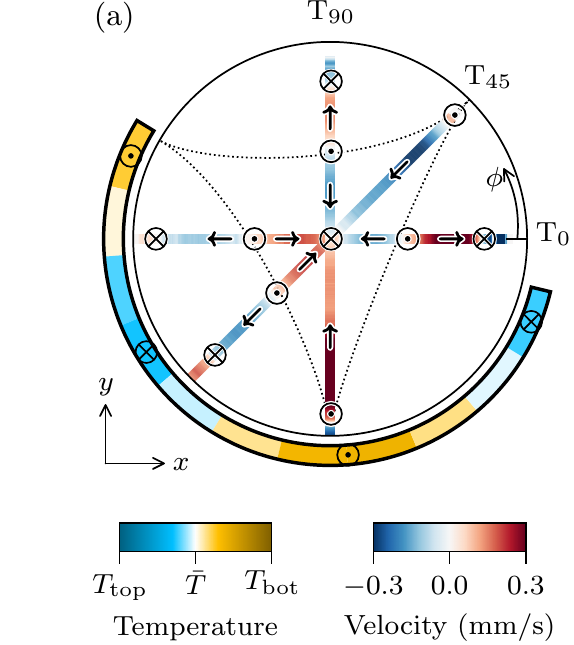}
\hspace{1em}
\includegraphics{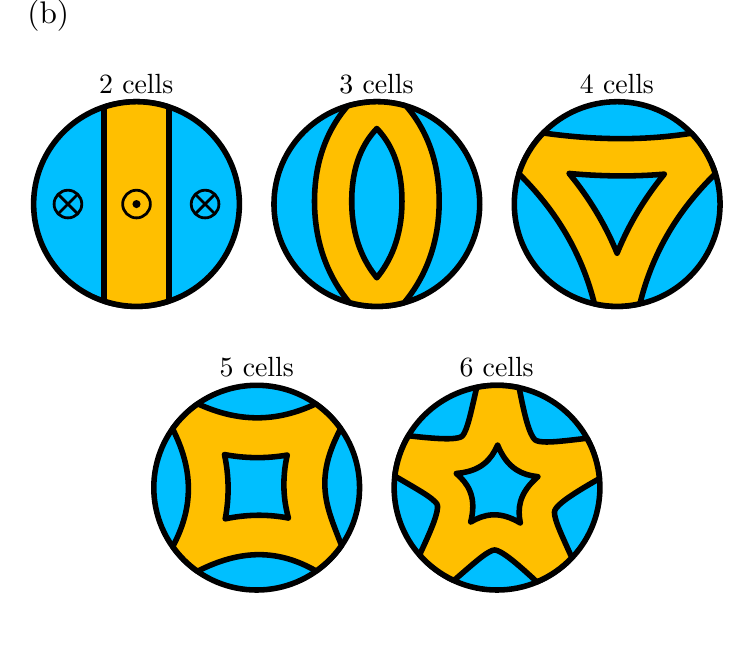}
\caption{%
  Identification of flow patterns in the cellular regime.
  (a)~Radial velocity profiles near the top plate and temperature profile at mid-height for $\Ra = 4.2 \times 10^6$ and $\Ha = 592$, averaged over $50 \tau_\mathrm{ff}$ (corresponding to figure~\ref{fig:Tmid_Ha}(d)).
  Velocity and temperature are colour-coded and shown at their position of measurement.
  Arrows indicate the direction of the flow.
  Positions of deduced up- and down-flows are marked by~$\odot$ and~$\otimes$, respectively.
  The dotted line shows the connection of upwelling flows separating regions of downwelling fluid.
  (b)~Schematic illustrations of the flow patterns in the cellular regime.
  The areas of up- and down-flow (orange and blue, respectively) over the horizontal cross-section at mid-height are illustrated.
  The flow in~(a) corresponds to the \textsl{4 cells} pattern.}
\label{fig:regime_cellular}
\end{figure}

To deduce the structure of the inner flow, the UDV velocity data is analysed.
Figure~\ref{fig:regime_cellular}(a) displays the time-averaged temperature profile from figure~\ref{fig:Tmid_Ha}(d) and the corresponding radial velocity data of UDV sensors T\textsubscript{0}, T\textsubscript{45} and T\textsubscript{90} near the top plate at their respective measurement position.
The directions of the flow are indicated by arrows.
We assume the absence of a significant azimuthal flow due to the axis-symmetry of the system which requires the presence of up- and downwelling fluid at the points of diverging and converging radial flows, respectively.
This is most noticeable in the centre of the cell, where the velocities of all three sensors converge into a down-flow along the central axis of the cell.
The deduced vertical flows are marked by~$\otimes$ and~$\odot$ in figure~\ref{fig:regime_cellular}(a) and are consistent with the three-fold symmetry from the temperature data at mid-height.
Small reversals of the velocity direction near the side walls indicate corner vortices.
It is noticeable, that the areas of down-flow (three along the circumference and one in the centre) are always separated by up-flows.
Connecting the positions of these upwellings gives a triangular shape (dotted line in figure~\ref{fig:regime_cellular}(a)).
The flow structure thus consists of four convection cells with down-streaming fluid in their centre and up-streaming fluid at the boundaries.
The velocity data of the other seven UDV sensors are consistent with this structure.

Using the above approach, five different flow patterns could be detected in the cellular regime consisting of two to six convection cells.
Their basic structures are displayed in figure~\ref{fig:regime_cellular}(b) by illustrating the vertical flow direction over the horizontal cross-section at mid-height.
They exhibit two-fold up to five-fold azimuthal symmetries.
Patterns with inverted flow directions occurred as well, i.e., up-flow in the centre of the convection cells and down-flow at the boundaries in between.
The pattern that is selected at a specific parameter combination $(\Ra,\Ha)$ is likely to depend on the time history of the experiment such as the speed at which the temperature difference is increased or decreased.
A similar observation in parameter regions of regime transitions was reported by \citet{Zhong1991} for rotating Rayleigh-B\'{e}nard convection. 
The classification of different patterns in the $(\Ra,\Ha)$ parameter space is detailed in section~\ref{sec:regimemap}.

The transition boundary from the LSC to the cellular flow regime (in short cellular regime) is not sharply defined.
At the crossover between both regimes, the large-scale flow may be highly transient and switch between a one- and multiple-roll structures intermittently.
In other instances, the flow could start in the cellular regime and return back into a one-roll LSC structure.
This is another indicator for the dependence of the flow state on the particular pathway through the $\Ra$--$\Ha$ parameter plane -- a typical hysteresis effect.

\begin{figure}
\centering
\includegraphics{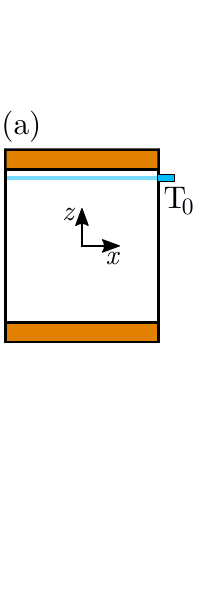}
\includegraphics{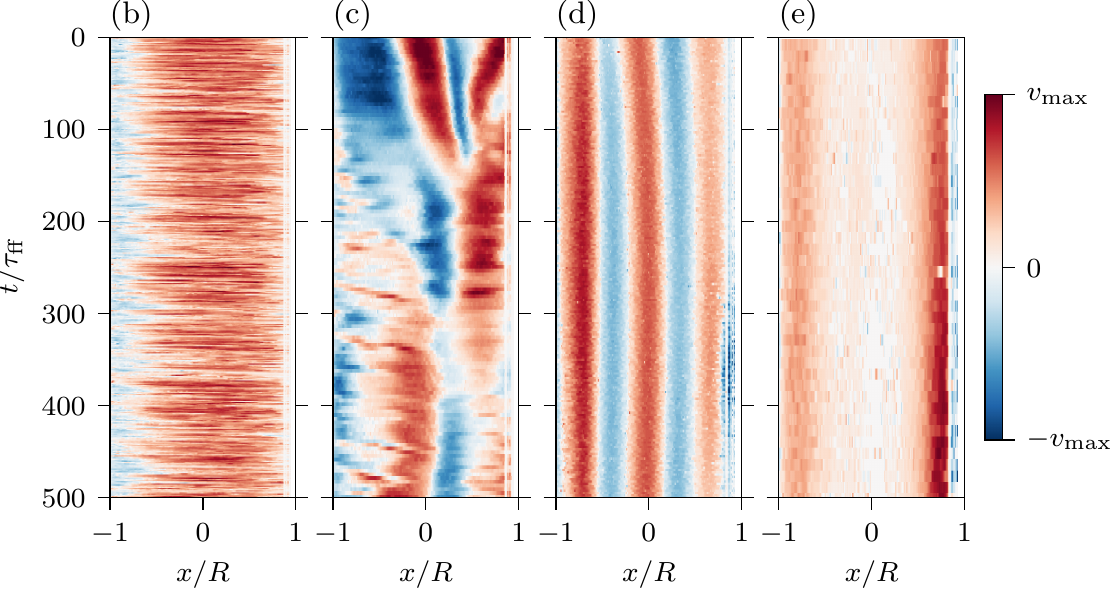}
\caption{%
  Velocity profiles $v_x(x,t)$ of the UDV sensor T\textsubscript{0} at $\Ra = 10^7$ ($\tau_\mathrm{ff} = 5.4$\,s) and different~$\Ha$.
  (a)~Illustration of the sensor position and measurement line (see also figure~\ref{fig:setup}(b)).
  (b)~LSC regime at $\Ha = 66$ with $v_\mathrm{max} = 20$\,mm\,s$^{-1}$.
  (c)~Transient cellular regime at $\Ha = 460$ with $v_\mathrm{max} = 2$\,mm\,s$^{-1}$.
  (d)~Stable cellular regime at $\Ha = 855$ with $v_\mathrm{max} = 1$\,mm\,s$^{-1}$.
  (e)~Wall mode regime at $\Ha = 1052$ with $v_\mathrm{max} = 0.3$\,mm\,s$^{-1}$.}
\label{fig:velo_Ha}
\end{figure}

This dynamic behaviour continues into the cellular regime.
Figure~\ref{fig:velo_Ha} displays the velocity profiles $v_x(x,t)$ of UDV sensor T\textsubscript{0} near the top plate (figure~\ref{fig:velo_Ha}(a)) for measurements at $\Ra=10^7$ and different $\Ha$.
The artefacts of the data near $x/R=1$ are the result of an inaccessible zone close to the UDV sensor due to the ringing of its piezoelectric transducer.
During the measurement at $\Ha = 66$ in figure~\ref{fig:velo_Ha}(b) the system is in the LSC regime.
The convection roll is mainly oriented in the $x$-$z$-plane; thus T\textsubscript{0} detects a positive $x$-velocity over the whole cell diameter.
Only a small area close to $x/R = -1$ shows negative velocities, indicating a recirculation vortex.
For $\Ha = 460$ in figure~\ref{fig:velo_Ha}(c), the flow structure has transformed into the cellular regime with multiple changes of the velocity directions along the measurement line.
The flow displays a highly transient behaviour without a clearly defined cell pattern.
These fluctuations are continually suppressed with increasing $\Ha$.
At $\Ha = 855$ (figure~\ref{fig:velo_Ha}(d)), the flow pattern has stabilised and the velocity field is quasisteady.
Any variation of the flow proceeds now on much larger time scales than in the transitional regime.
Here, the free-fall time~$\tau_\mathrm{ff}$ loses its relevance as a time scale due to the strong magnetic damping.
These exemplary velocity data illustrate that the flow structures of the cellular regime can exhibit complex dynamical properties.
Along with the history-dependency of the pattern selection, a more detailed analysis of this behaviour has to be addressed in future experiments.

\subsection{The wall mode regime}
\label{sec:wallmode}
\begin{figure}
\centering
\includegraphics{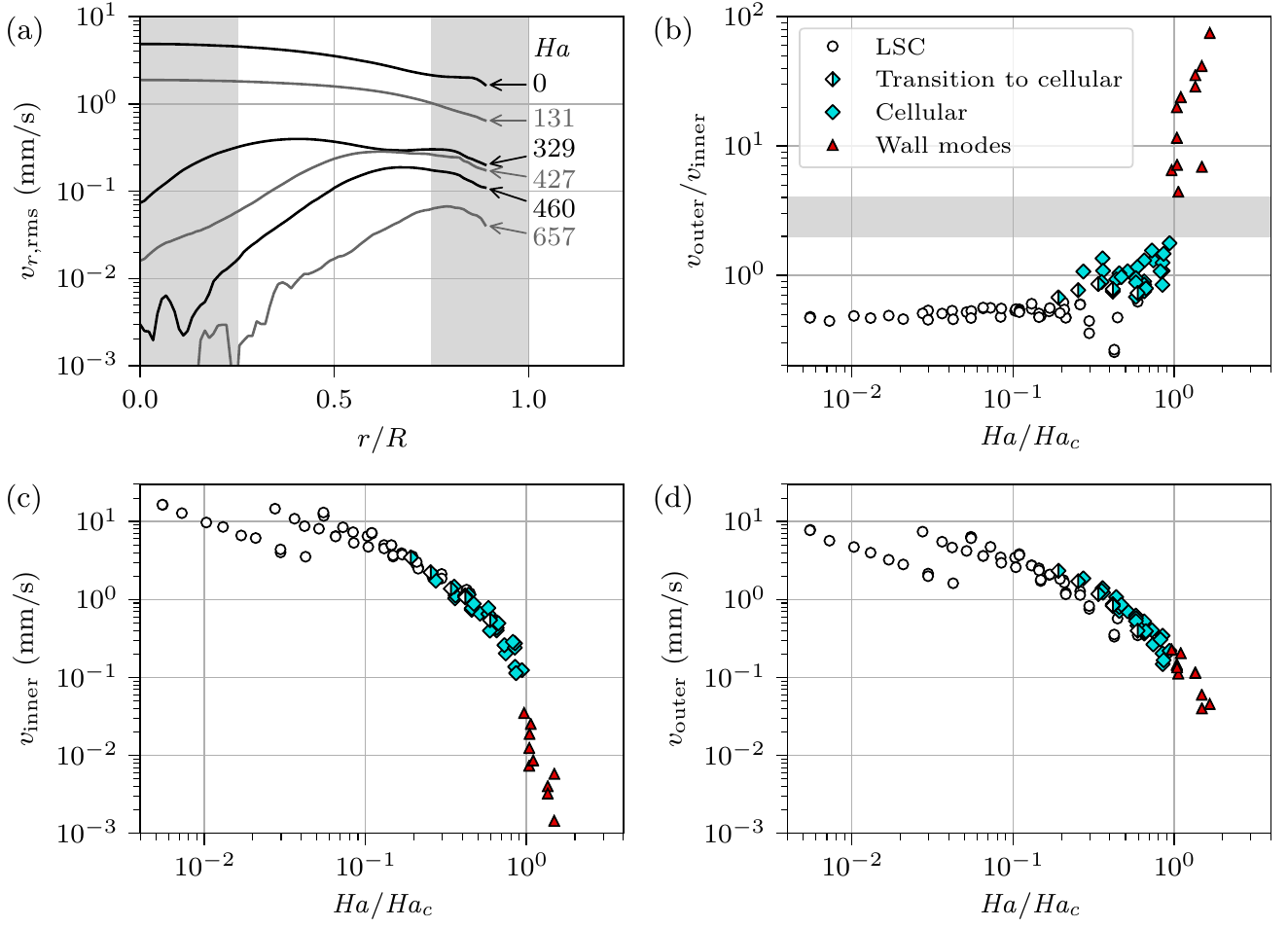}
\caption{%
  (a)~Profiles of the radial root mean square velocity component~$v_{r,\mathrm{rms}}$ with increasing $\Ha$ and for $\Ra=2.1\times10^6$.
  The critical Hartmann number for the linear instability threshold is in this case $\Ha_\mathrm{c} = 443$.
  The grey areas mark the intervals with respect to $r$ which have been used for calculating $v_\mathrm{outer}$ and $v_\mathrm{inner}$ which are defined by eqn.~\eqref{eq:inout}.
  (b)~Ratio of the outer and inner velocity $v_\mathrm{outer}/v_\mathrm{inner}$.
  Flow regimes are indicated by different symbols.
  The grey area marks the gap between the cellular and wall mode regimes.
  (c,d)~Replot of the inner and outer velocities for all Rayleigh and Hartmann numbers.
  Symbols correspond to~(b).
  Jumps in the data are due to the different $\Ra$ they are recorded at.}
\label{fig:regime_wallmode}
\end{figure}

The final regime detected in the present experiments appears for the highest $\Ha$ and lowest $\Ra$.
The flow field in figure~\ref{fig:velo_Ha}(e) at $\Ra = 10^7$ and $\Ha=1052$ shows a distinctly different profile than at lower $\Ha=855$ in figure~\ref{fig:velo_Ha}(d).
The velocity magnitude in the cell centre ($x/R=0$) is much smaller compared to the areas near the side wall ($x/R=\pm1$).
This change in the flow field cannot be detected by the thermocouple array at mid-height.
Figure~\ref{fig:Tmid_Ha}(e) shows the temperature profile corresponding to the first $50\tau_\mathrm{ff}$ of figure~\ref{fig:velo_Ha}(e).
It shows no significant difference towards the cellular regime (figure~\ref{fig:Tmid_Ha}(d)).

To identify this new regime, we consider the radial velocity profiles $v_r(r,t)$ recorded by the UDV sensors T\textsubscript{0}, T\textsubscript{45}, T\textsubscript{90}, M\textsubscript{0}, M\textsubscript{90}, B\textsubscript{0}, B\textsubscript{45} and B\textsubscript{90}.
In order to eliminate noise, a median filter is applied to the raw velocity data with a size of 5\,mm along the radial direction and over five consecutive profiles.
Subsequently, the root mean square (rms) of these profiles is taken with respect to time and over all sensors to obtain a radial profile of the radial velocity intensity~$v_{r,\mathrm{rms}}(r)$.
These profiles are plotted for $\Ra = 2.1 \times 10^6$ and different $\Ha$ in figure~\ref{fig:regime_wallmode}(a).
The overall magnitude of the velocity decreases with increasing $\Ha$, an observation which is expected.
In the LSC regime at $\Ha= 0$ and 131, the rms velocity magnitude is largest in the centre of the cell ($r/R = 0$) and decreases towards the side-wall ($r/R = 1$).
The profile in the cellular regime, e.g., at $\Ha = 329$, shows a more evenly distributed magnitude that decreases towards the centre.
For higher magnetic fields at $\Ha = 427$, 460 or 657, the rms magnitude in the centre of the cylinder is drastically decreased compared to the near-wall region.
This is the so-called wall mode regime.
The variations of these profiles near the centre $r=0$ are due noise and measurement artefacts, which appear at low velocity magnitudes of the order of $10^{-2}$\,mm/s and below.

Averaging the $v_{r,\mathrm{rms}}$ profiles over the inner and outer 25\,\% of the radius defines a typical inner and outer velocity, respectively.
Both quantities are given by
\begin{equation}
v_\mathrm{inner} 
  = \langle v_{r,\mathrm{rms}}(r)\rangle_{r \le 0.25 R} \,, \qquad
v_\mathrm{outer} 
  = \langle v_{r,\mathrm{rms}}(r)\rangle_{r \ge 0.75 R} \,.
\label{eq:inout}
\end{equation}
The ratio $v_\mathrm{outer} / v_\mathrm{inner}$ is displayed in figure~\ref{fig:regime_wallmode}(b) over $\Ha$ normalized by $\Ha_\mathrm{c}$ at the Rayleigh number of the respective measurement.
The amplitudes of both velocities are shown in panels (c) and (d) of this figure.
In the LSC regime the ratio is $\sim 0.5$ and increases up to 2 in the cellular regime.
At $\Ha/\Ha_\mathrm{c} = 1$, there is a large gap in the data, signifying a sudden strong decrease of $v_\mathrm{inner}$ compared to $v_\mathrm{outer}$.
This means the flow is suppressed in the cell centre and pushed towards the side walls.
These wall modes were also found in recent direct numerical simulations by \citet{Liu2018}.
The critical Hartmann number $\Ha_\mathrm{c}$ marks the onset of convection for an infinite fluid layer without side-walls.
The sharp transition to wall modes at $\Ha = \Ha_\mathrm{c}$ shows, that the flow in the fluid bulk behaves as in an unbounded layer.
The remaining flow for $\Ha > \Ha_\mathrm{c}$ is a result of the electrically insulating side-walls, which reduce the effect of the magnetic damping~\cite{Houchens2002}.
The thermocouple array at the side wall of our experiment detects multiple up- and down-flows as can be seen in figure~\ref{fig:Tmid_Ha}(e).
Similar to the cellular regime, the order of azimuthal symmetry depends on the parameter combination $(\Ra, \Ha)$. 
More information can be found in the following section~\ref{sec:regimemap}.
Future measurements will address the history-dependence of the pattern selection in detail.

\subsection{Regime map}
\label{sec:regimemap}

\begin{figure}
\centering
\includegraphics{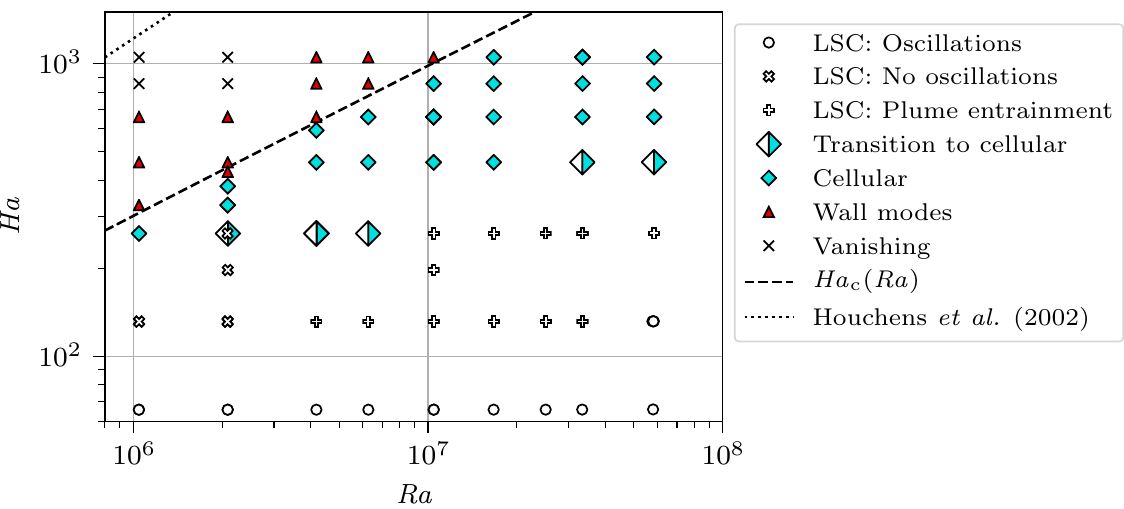}
\caption{%
  Regime map over the $(\Ra, \Ha)$ parameter space.
  Measurements for $\Ha < 60$ (not shown) belong to the LSC regime with regular oscillations.
  The lines indicate the theoretical onsets of convection for an infinite fluid layer $\Ha_\mathrm{c}$ by \citet{Chandrasekhar1961} and a cylindrical cell of $\Gamma=1$ by \citet{Houchens2002}.}
\label{fig:regimemap}
\end{figure}

Figure~\ref{fig:regimemap} summarizes the detected flow regimes over the $(\Ra, \Ha)$ parameter space.
All measurements $\Ha < 100$ display a LSC structure with regular sloshing and torsion oscillations (denoted as \textsl{LSC: Oscillations}).
Only for the highest Rayleigh number of $\Ra = 6 \times 10^7$ this regime is maintained up to Hartmann number of $\Ha=131$.
For higher $\Ha$, the regular oscillations are suppressed but the LSC is still retained (\textsl{LSC: No oscillations}).
Above a Rayleigh number of $\Ra > 3 \times 10^6$, the flow additionally displays the plume entrainment pattern (\textsl{LSC: Plume entrainment}).
The subsequent breakdown of the LSC is not uniquely defined and multiple measurements display properties of the LSC and the cellular regime (\textsl{Transition to cellular}).
The critical Hartmann number~$\Ha_\mathrm{c}$ for an infinite fluid layer~\cite{Chandrasekhar1961} clearly marks the onset of the wall mode regime.
Only one measurement at $\Ra = 2.1 \times 10^6$ and $\Ha/\Ha_\mathrm{c} = 0.97$ displays properties of the wall mode regime for $\Ha < \Ha_\mathrm{c}$.
The measurement closest to $\Ha_\mathrm{c}$ in the cellular regime is at a Rayleigh number  $\Ra = 4.2\times 10^6$ and for $\Ha/\Ha_\mathrm{c} = 0.94$.
At the highest Hartmann numbers $\Ha > 800$ and for Rayleigh numbers $\Ra < 3 \times 10^6$, the temperature and velocity measurements cannot detect a distinctive flow structure anymore (\textsl{Vanishing}), i.e., the recorded values are below the measurement resolution.
This does not necessarily imply that convection is completely suppressed by the magnetic field.
A flow very close to the side-wall is outside the measurement capabilities of the UDV sensors and could be too slow to induce significant temperature variations at mid-height.
\citet{Houchens2002} performed a linear stability analysis for magnetoconvection in a cylindrical cell of aspect ratio $\Gamma = 1$.
They found the onset at $\Ha = 0.122 \Ra^{2/3}$ for $\Ra \gtrsim 10^5$ (dotted line in figure~\ref{fig:regimemap}) which our experiments cannot cross.
In other words, electrically insulating side walls destabilize the magnetoconvection flow, a phenomenon that is similar to rotating convection~\cite{Zhong1991}.

Our experiments satisfy the condition that $\kappa<(\mu_0\sigma)^{-1}$ or that of $\Pr \gg \Pm$ which suppresses the overstability and thus the appearance of an oscillatory regime~\cite{Chandrasekhar1961}. 
Simulations of magnetoconvection by \citet{Yan2019} at $\Pr = 1$ in Cartesian cells with periodic horizontal boundary conditions also reveal a cellular and a turbulent regime, the latter of which corresponds to the LSC regime in the present study.
In the crossover between both regimes, a regime with narrow columnar up- and downflows is detected which are not destroyed by overstability. 
This regime existed also in their simulations at $\Pr=0.025$ for $\Ha \ge 1414$. 
We cannot detect these structures in our experiment and point to the different geometry and velocity boundary conditions, stress-free in combination with periodic conditions in the simulations versus no-slip at all walls in our experiment, as one possible reason for this discrepancy.

\begin{figure}
\centering
\includegraphics{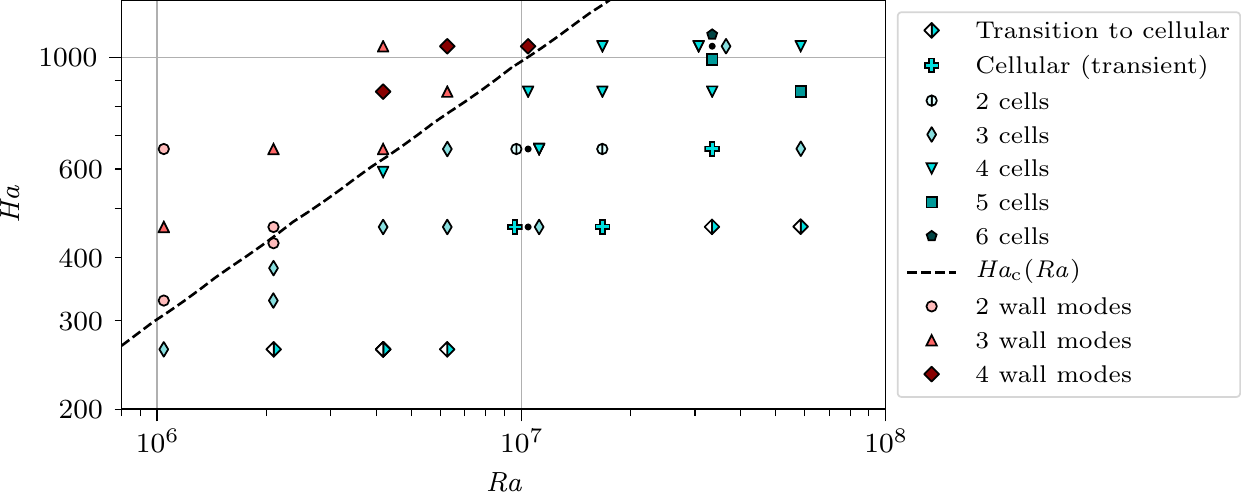}
\caption{%
  Distribution of flow patterns, represented by the marker shape, in the cellular and wall mode regime over the $(\Ra, \Ha)$ parameter space.
  The dashed line indicates the theoretical onsets of convection for an infinite fluid layer $\Ha_\mathrm{c}$~\cite{Chandrasekhar1961}.
  If multiple patterns have been observed at the same parameter combination, the respective markers are clustered around a black dot which indicates the actual measurement location.
  Measurements in the cellular regime that could not be properly identified due to strong fluctuations are assigned to the \textsl{Cellular (transient)}.}
\label{fig:regimemap_cell+wall}
\end{figure}

Figure~\ref{fig:regimemap_cell+wall} replots the regime map and highlights the different flow patterns of the cellular and wall mode regimes. 
The patterns in the cellular regime are named according to figure~\ref{fig:regime_cellular}(b).
The number of wall modes is determined by means the temperature array at mid-height.
For example, the \textsl{3 wall modes} pattern consist of three up- and three down-flows over the whole circumference.
Patterns of type \textsl{Transition to cellular} and \textsl{Cellular (transient)} without a specific cell number are all situated at the low-$\Ha$ boundary towards the LSC regime.
An overall trend to a higher symmetry and cell number with increasing~$\Ra$ is visible.
A refinement of this map with a finer sampling of $(\Ra, \Ha)$ or an inclusion of the dependence of pattern selection on the history of the temperature variation (hysteresis effects) is beyond the scope of the present work. 
Our presented regime map in figure~\ref{fig:regimemap} provides a general overview over the various flow structures occurring in liquid metal magnetoconvection.

In a number of measurements the flow experienced a constant azimuthal drift of up to three full turns per hour, an effect which occurred for several parameter combinations $(\Ra, \Ha)$. 
An effect of the Coriolis force of the earth, as discussed by \citet{Brown2006} for RBC without magnetic fields, can be excluded as the reason of this drift since it only occur for $\Ha > 0$.
Additionally, a reversal of the vertical external magnetic field polarity caused a reversal of the drift direction. 
The Lorentz forces induced by the interaction of the velocity field and the external magnetic field are consequently not the reason for the drift since they are invariant under the inversion of the magnetic field~\cite{Davidson2001}. 
However, the magnetic field could be responsible for an additional force on the flow by the thermoelectric effect, which results from non-constant temperature differences between copper and liquid metal along the flow direction. 
These temperature differences are caused by the large-scale circulation and generate a thermoelectric current in radial direction.
Due to the vertical magnetic field, this thermoelectric current induces a Lorentz force which acts on the liquid in azimuthal direction. 
To prevent this effect, the copper plates would have to be covered with an electrically insulating layer to electrically separate the copper from the liquid metal.

\section{Global transport properties}
\label{sec:transport}
\begin{figure}
\centering
\includegraphics{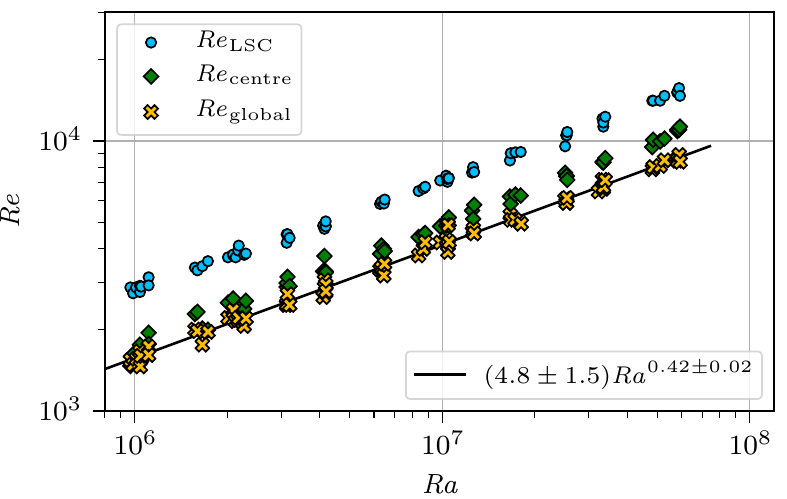}
\caption{%
  Global Reynolds number~$\Re_\mathrm{global}$ vs.\ Rayleigh number~$\Ra$ at $\Ha=0$.
  Values of $\Re_\mathrm{LSC}$ (based on the LSC velocity at the plates) and $\Re_\mathrm{centre}$ (based on the velocity fluctuations in the cell centre) are replotted from \citet{Zurner2019}.}
\label{fig:Reglobal}
\end{figure}

The global heat transport is characterized by the Nusselt number $\Nu = \dot Q / \dot Q_\mathrm{cond}$, where $\dot Q_\mathrm{cond} = \lambda\pi R^2 \Delta T / H$ is the purely conductive heat flux with the heat conductivity~$\lambda$.
The Reynolds number~$\Re = UL/\nu$ which quantifies the momentum transport in the anisotropic and inhomogeneous convection flow does not have a clear definition since its magnitude depends on the choice of the characteristic velocity $U$, the measurement position and method as discussed in \citet{Zurner2019}.
An adaption of the definition of $U$ to the large variety of flow structures in the different regimes described in section~\ref{sec:large-scale} introduces biases.
We want to provide an unique definition of $\Re$ that applies for the whole parameter space.
Numerical simulations typically calculate $U$ as a rms-average over the whole fluid volume.
Following this approach, we calculate a global velocity scale $v_\mathrm{global}$ by taking the rms-average of all velocity data recorded by the ten UDV sensors (over position and time).
The resulting Reynolds number $\Re_\mathrm{global} = v_\mathrm{global}H/\nu$ is plotted for $\Ha=0$ in figure~\ref{fig:Reglobal}.
For comparison, the Reynolds numbers $\Re_\mathrm{LSC}$ and $\Re_\mathrm{centre}$, based on the LSC velocity near the plates and the velocity fluctuations in the cell centre, are replotted from \citet{Zurner2019}.
The magnitude of $\Re_\mathrm{global}$ is lower than $\Re_\mathrm{LSC}$ and $\Re_\mathrm{centre}$, since the UDV sensors probe many low-velocity areas of the flow.
A power law fit results in a scaling of $\Re_\mathrm{global} \simeq (4.8\pm1.5) \Ra^{0.42\pm0.02}$ with the same exponent as $\Re_\mathrm{LSC} \simeq (8.0 \pm 4.4) \Ra^{0.42\pm0.03}$~\cite{Zurner2019}, showing that it is dominated by the high-velocity areas of the flow.
It should be noted, that this definition of the Reynolds number $\Re_\mathrm{global}$ can still introduce some biases towards different flow states.
The reason is that UDV sensors can probe a limited number of regions and velocity components of the fluid only.
Any large-scale flow which is predominantly located in those areas and aligned with some of the crossing beam lines will intrinsically give a stronger signal than a flow of the same intensity which is centred at different regions.
Our results will be discussed in light of this bias.

For the following, the Reynolds number and the convective part of the Nusselt number will always be normalized by their reference values at $Ha=0$ which leads to 
\begin{equation}
\tilde \Nu
  = \frac{\Nu(\Ra, \Ha)-1}{\Nu(\Ra, 0)-1} \,, \qquad\qquad    
\tilde \Re_\mathrm{global}
  = \frac{\Re_\mathrm{global}(\Ra, \Ha)}{\Re_\mathrm{global}(\Ra, 0)} \,.
\end{equation}
\begin{figure}
\centering
\includegraphics{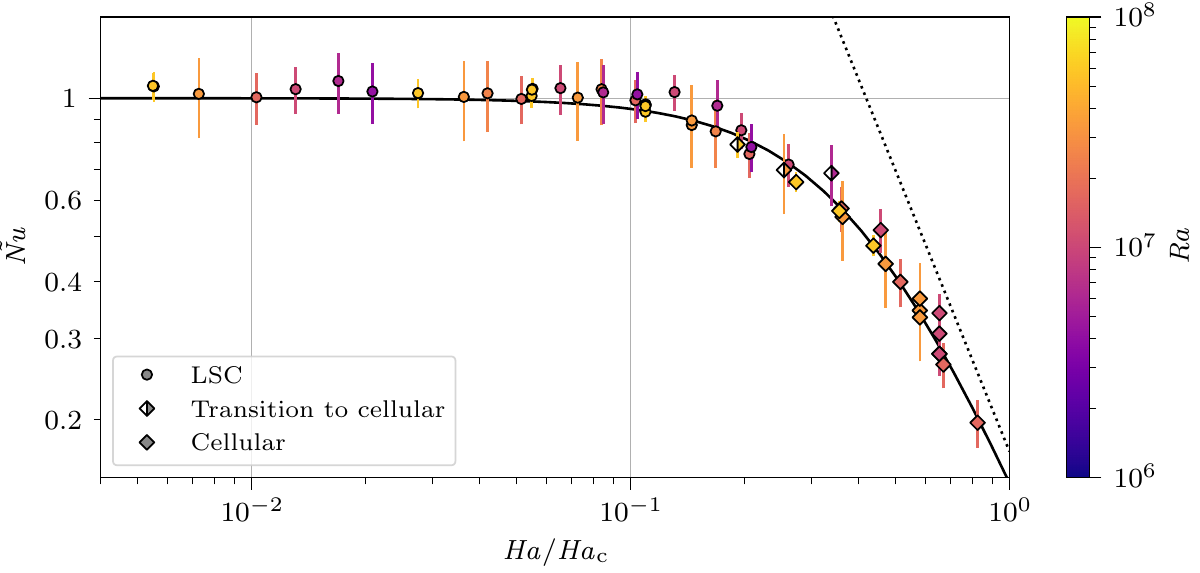}
\caption{%
  Normalized Nusselt number~$\tilde\Nu$ vs.\ Hartmann number~$\Ha$, normalized by $\Ha_\mathrm{c}$.
  Flow regimes and Rayleigh numbers are indicated by differently shaped and coloured symbols, respectively.
  The fit of \eqref{eq:Nu_fit} is plotted as solid line with parameters $\chi_1 = 5.9$ and $\gamma_1 = 2.03$.
  Its asymptote for high $\Ha$ is shown as dotted line.}
\label{fig:Nu(Ha)}
\end{figure}

The normalized Nusselt number~$\tilde\Nu$ is plotted versus $\Ha/\Ha_\mathrm{c}$ in figure~\ref{fig:Nu(Ha)}.
With this normalization, as in \citet{Cioni2000}, the data collapse to a single curve for all Rayleigh numbers.
For nearly the whole LSC regime $\tilde \Nu$ retains its original value from $\Ha=0$.
Only close to the transition into the cellular regime, $\tilde\Nu$ starts to drop to values of $\sim 80\,\%$.
Once the cellular regime is reached, the Nusselt number drops significantly with increasing Hartmann number.
The resulting dependence of $\tilde\Nu$ can be described by
\begin{equation}
\label{eq:Nu_fit}
\tilde\Nu \simeq 
  \left[1 + \chi_1\left(\frac{\Ha}{\Ha_\mathrm{c}}\right)^{\gamma_1}
  \right]^{-1} \,.
\end{equation}
A least-squares fit to the data with the parameters $\chi_1 = 5.9 \pm 0.3$ and $\gamma_1 = 2.03 \pm 0.06$ is plotted as a solid line in figure~\ref{fig:Nu(Ha)}.
For large $\Ha$, \eqref{eq:Nu_fit} becomes the power law $\tilde\Nu \simeq 0.17(\Ha/\Ha_\mathrm{c})^{-2.03}$ (dotted line).
This asymptote is, however, not reached before the transition to the wall mode regime at $\Ha/\Ha_\mathrm{c} = 1$.
Lacking a sufficient amount of data in this regime, it is not clear whether the asymptote is reached eventually or whether the scaling changes for $\Ha > \Ha_\mathrm{c}$, a point which has to be left open for now.

\begin{figure}
\centering
\includegraphics{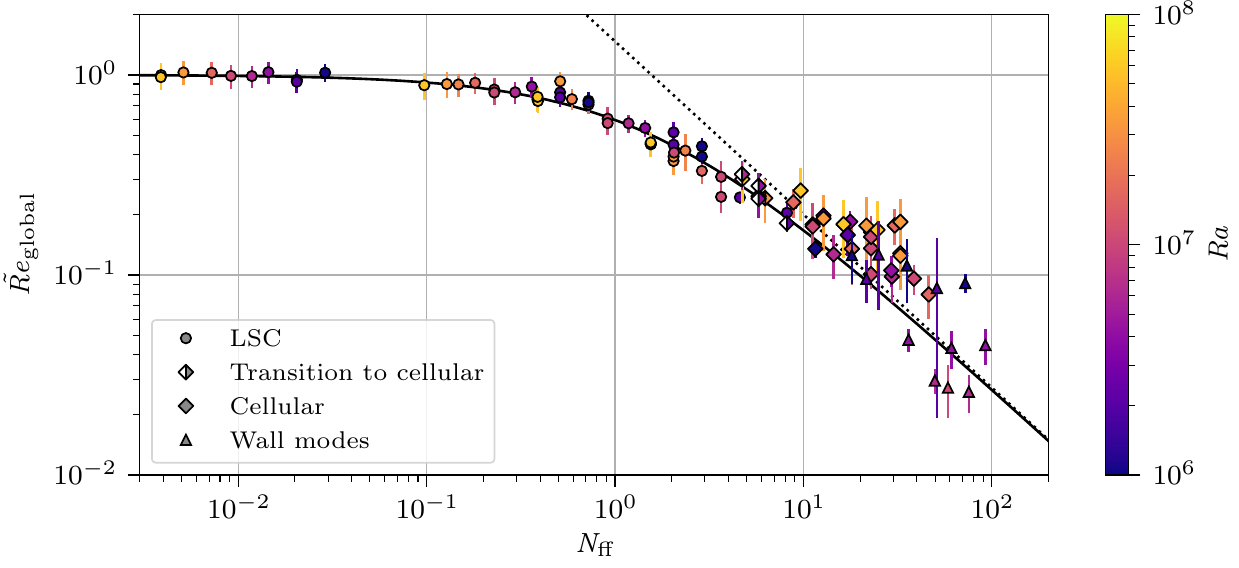}
\caption{%
  Normalized Reynolds number~$\tilde\Re_\mathrm{global}$ vs.\ the free-fall interaction parameter~$N_\mathrm{ff}$.
  Flow regimes and Rayleigh numbers are indicated by differently shaped and coloured symbols, respectively.
  The fit of~\eqref{eq:Re_fit} is plotted as solid line with parameters $\chi_2 = 0.68$ and $\gamma_2 = 0.87$.
  Its asymptote for high $\N_\mathrm{ff}$ is shown as dotted line.}
\label{fig:Re(Nff)}
\end{figure}

The Reynolds number displays a similar progression as the Nusselt number.
However, the $\tilde\Re_\mathrm{global}$ data do not collapse over $\Ha/\Ha_\mathrm{c}$.
To find a better dependency, the function~\eqref{eq:Nu_fit} is modified to $[1+\chi_2' (\Ha / \Ha_\mathrm{c}^\beta)^{\gamma_2'}]^{-1}$ with an additional fit parameter~$\beta$.
A fit to $\tilde\Re_\mathrm{global}$ results in $\beta = 0.47 \pm 0.04 \approx 0.5$ (as well as $\chi_2' = 0.04\pm0.02$ and $\gamma_2' = 1.74 \pm 0.05$).
This suggests, that the Reynolds number scales with $\Ha / \sqrt{\Ha_\mathrm{c}}$.
Using $\Ha_\mathrm{c} = \sqrt{\Ra}/\pi$ in the Chandrasekhar limit and introducing the free-fall Reynolds number $\Re_\mathrm{ff} = \sqrt{\Ra/\Pr}$ gives 
\begin{equation}
\label{eq:Re_argument}
\frac{\Ha}{\sqrt{\Ha_\mathrm{c}}}
  = \sqrt{\frac{\Ha^2}{\Re_\mathrm{ff}} \frac{\pi}{\sqrt{\Pr}}}
  = \sqrt{N_\mathrm{ff} \frac{\pi}{\sqrt{\Pr}}}  \,.
\end{equation}
Here, the free-fall interaction parameter~$N_\mathrm{ff} = \Ha^2/\Re_\mathrm{ff}$ quantifies the ratio of magnetic and buoyant forcing in the fluid.
Figure~\ref{fig:Re(Nff)} shows that the $\tilde\Re_\mathrm{global}$ data indeed collapse onto one curve when plotted over $N_\mathrm{ff}$.
The Reynolds number continuously decreases over the LSC regime range for  $\N_\mathrm{ff} \gtrsim 0.1$, reaching values of $\tilde\Re_\mathrm{global} \sim 30\,\%$ at the crossover to the cellular regime.
In the cellular and wall mode regimes, the data are more strongly scattered than in the LSC regime, because of the different flow configurations which are probed locally by the UDV sensors.
However, a continuous decrease of $\tilde\Re_\mathrm{global}$ is still observed.
A fit of the function
\begin{equation}
\label{eq:Re_fit}
\tilde\Re_\mathrm{global} \simeq 
  \Bigl[1 + \chi_2 N_\mathrm{ff}^{\gamma_2} \Bigr]^{-1} 
\end{equation}
to the data gives $\chi_2 = 0.68 \pm 0.04$ and $\gamma_2 = 0.87 \pm 0.03$, shown in figure~\ref{fig:Re(Nff)} as a solid line.
It should be noted that $\chi_2$ could in general be a function of $\Pr$ (see eqn.~\eqref{eq:Re_argument}).
Here, this is not relevant since $\Pr \approx \mathrm{const}$ (the $\Pr$-dependence of $\tilde\Re$ is discussed at the end of this section).
The asymptotic behaviour of \eqref{eq:Re_fit} in the cellular and wall modes regime is the power law $\tilde\Re_\mathrm{global} \simeq 1.47 N_\mathrm{ff}^{-0.87}$ for large $\N_\mathrm{ff} \gtrsim 10$ (see dotted line in figure~\ref{fig:Re(Nff)}).
The transition from $\tilde\Re\simeq 1$ for small $\N_\mathrm{ff}$ to the high-$\N_\mathrm{ff}$ asymptote is centred around $\N_\mathrm{ff} \sim 1$ with $1.47 \N_\mathrm{ff}^{\,-0.87} = 1$ at $\N_\mathrm{ff} \approx 1.6$

\begin{figure}
\centering
\includegraphics{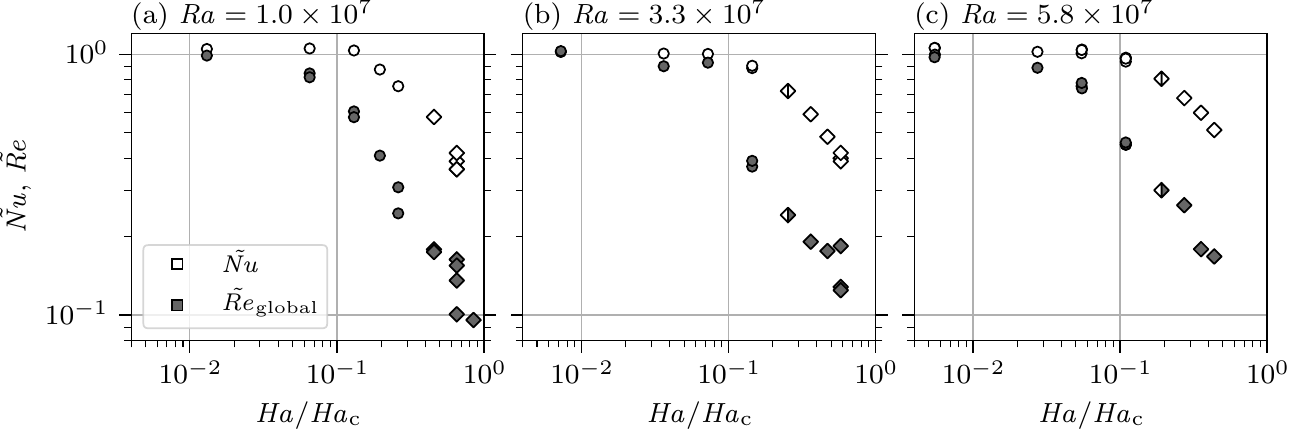}
\caption{%
  Comparison of $\tilde\Nu$ (open symbols) and $\tilde\Re_\mathrm{global}$ (grey symbols) over $\Ha/\Ha_\mathrm{c}$ for three selected $\Ra$.
  Flow regimes are indicated by the same symbol shapes as in figures~\ref{fig:Nu(Ha)} and~\ref{fig:Re(Nff)}.}
\label{fig:compare_NuRe}
\end{figure}

$\tilde\Nu$ and $\tilde\Re$ can be compared by their regime (marker shape) in figures~\ref{fig:Nu(Ha)} and~\ref{fig:Re(Nff)}. 
Additionally, figure~\ref{fig:compare_NuRe} compares the data for three selected $\Ra$ directly. 
It is remarkable that the global heat transport stays nearly constant up to the transition to the cellular regime while the global momentum transport decreases continually over the LSC regime.
This decoupling of transport properties from the average flow speed is also observed for other convection systems that include a stabilizing force, e.g.\ confined or rotating convection~\cite{Chong2017}.
The magnetic damping decelerates the flow, but more importantly, suppresses the turbulent oscillations and thus increases the coherence of the large-scale fluid motion.
This in turn improves the convective heat transport and compensates the effect of an overall slower flow magnitude.
In simulations of magnetoconvection at $\Pr\sim8$, \citet{Lim2019} even found an increase of $\Nu$ for an optimal $\Ha$.
Such an improvement of the transport properties is generally associated with the crossover of the viscous and the thermal boundary layer (BL)~\cite{Chong2017, Lim2019}.
For low-$\Pr$ convection, the viscous BL is much thinner than the thermal BL at $\Ha = 0$~\cite{Stevens2013, Scheel2016,Scheel2017}.
By increasing $\Ha$ the viscous BL thickness further shrinks as it has to be substituted by a Hartmann BL thickness, which is given by $\delta_B = H/\Ha$, whereas the thermal BL thickness $\delta_T = H/(2\Nu)$ grows~\cite{Yan2019}.
Consequently, a BL crossover is not expected in the present experiments.
Indeed, an associated increase of $\Nu$ with $\Ha$ cannot be observed in figure~\ref{fig:Nu(Ha)}.

The magnetic Reynolds number $\Rm = \Re\Pm$ is always significantly smaller than~1.
The highest Reynolds numbers are reached for $\Ha=0$, since $\Re$ decreases with increasing $\Ha$.
The maximal values of $\Re$ shown in figure~\ref{fig:Reglobal} do not exceed $2 \times 10^4$ (see also \citet{Zurner2019}).
With a magnetic Prandtl number $\Pm = 1.36 \times 10^{-6}$ this results to $\Rm < 2.72 \times 10^{-2}$.
Thus, the quasistatic approximation is applicable within the whole parameter space of the present experiment, i.e., the applied magnetic field is not significantly distorted by the convective flow~\cite{Davidson2001}.

\begin{figure}
\centering
\includegraphics{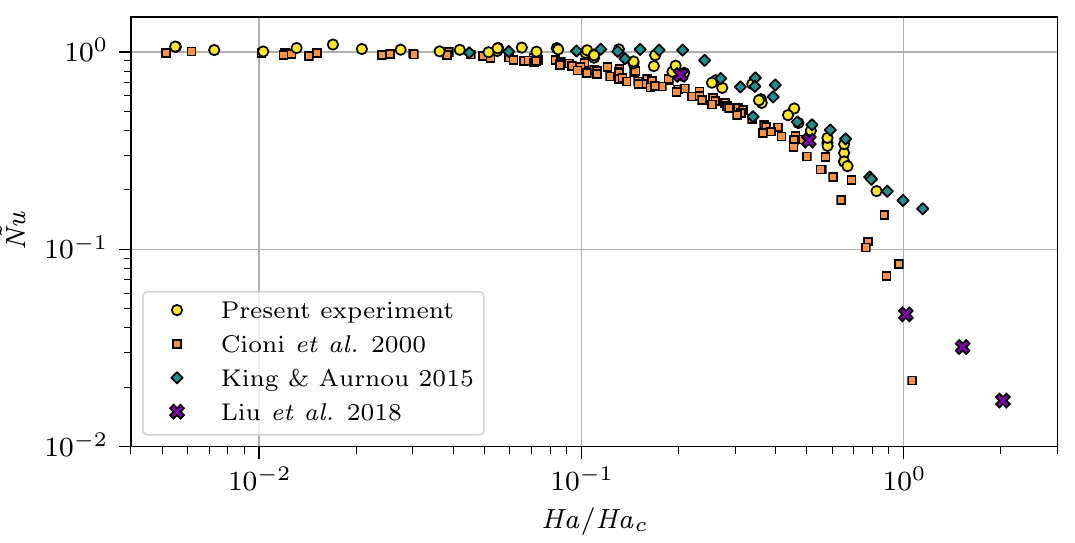}
\caption{%
  Comparison of the normalized Nusselt number data as a function of the Hartmann number with data at $\Pr = 0.025$ from experiments by \citet{Cioni2000} and \citet{King2015}, as well as direct numerical simulations by \citet{Liu2018}.
  The circles are the present data replotted from figure~\ref{fig:Nu(Ha)}.}
\label{fig:literature_Nu}
\end{figure}
\begin{figure}
\centering
\includegraphics{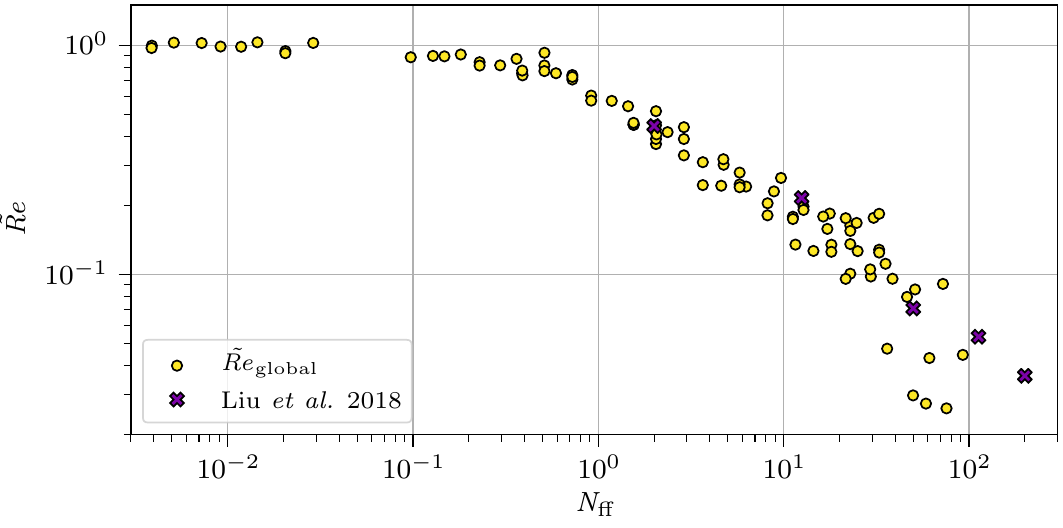}
\caption{%
  Comparison of normalized Reynolds number data as a function of the free-fall interaction parameter with direct numerical simulations by \citet{Liu2018} at $\Pr = 0.025$.
  The circles are the present data replotted from figure~\ref{fig:Re(Nff)}.
  The Reynolds numbers by \citet{Liu2018} are based on the root mean square of the velocity field over the whole fluid volume for $\Ra=10^7$.}
\label{fig:literature_Re}
\end{figure}
We now compare the results of the present experiments with further data from the literature.
Nusselt number measurements by \citet{Cioni2000} and \citet{King2015}, as well as simulations by \citet{Liu2018} (all at $\Pr=0.025$) are shown in figure~\ref{fig:literature_Nu}.
The data are plotted again over $\Ha/\Ha_\mathrm{c}$ and are normalized by their respective values at $\Ha=0$, which agree well with the present experiment~\cite{Zurner2019}.
\citet{Liu2018} find a Nusselt number of $Nu=9.75\pm0.05$ at $Ha=0$ and $\Ra=10^7$, which is consistent with our result $Nu=9.2\pm0.7$~\cite{Zurner2019}.
Simulations by \citet{Yan2019} at $\Pr=0.025$ agree with the data by \citet{Cioni2000}, but do not report results for $\Ha=0$ and are thus not shown here.
The data follow the same general progression as the present results.
The experiments deviate slightly from one another once $\tilde Nu$ drops off, even though every individual set of data collapses over its range of $Ra$.
For $\Ha > \Ha_\mathrm{c}$ the data disagree more strongly, giving a range of $0.02 < \tilde\Nu < 0.3$.
Since the wall modes depend on the side walls of the cell, one reason may be the different geometry: \citet{Liu2018} use a rectangular box of $\Gamma=4$, while the experiments by \citet{Cioni2000} and \citet{King2015} are conducted in cylinders of $\Gamma=1$.
The large deviation between the experiments is less clear and may be attributed to the difficulty of accurately measuring low $\Nu \sim 1$.

The authors are to the best of their knowledge not aware of any previous experiments that report Reynolds numbers for the magnetoconvection case.
Direct numerical simulations by \citet{Liu2018} and \citet{Yan2019} report such measurements for $\Pr = 0.025$.
The latter DNS cannot be compared since it reports data for $\Ha = 1414$ outside our range of $\Ha$ and is missing data at $\Ha=0$ needed for the normalization of $\tilde\Re$.
The former DNS is plotted over $\N_\mathrm{ff}$ along with $\tilde \Re_\mathrm{global}$ in figure~\ref{fig:literature_Re}.
The two data sets agree very well with one another, though a fit of \eqref{eq:Re_fit} to the simulation gives a somewhat smaller decrease ($\chi_2 = 0.78$, $\gamma_2 = 0.66$) than the present experimental results, which might be explained with the difference in geometry or the limited velocity data acquired by the UDV sensors.

\begin{figure}
\centering
\includegraphics{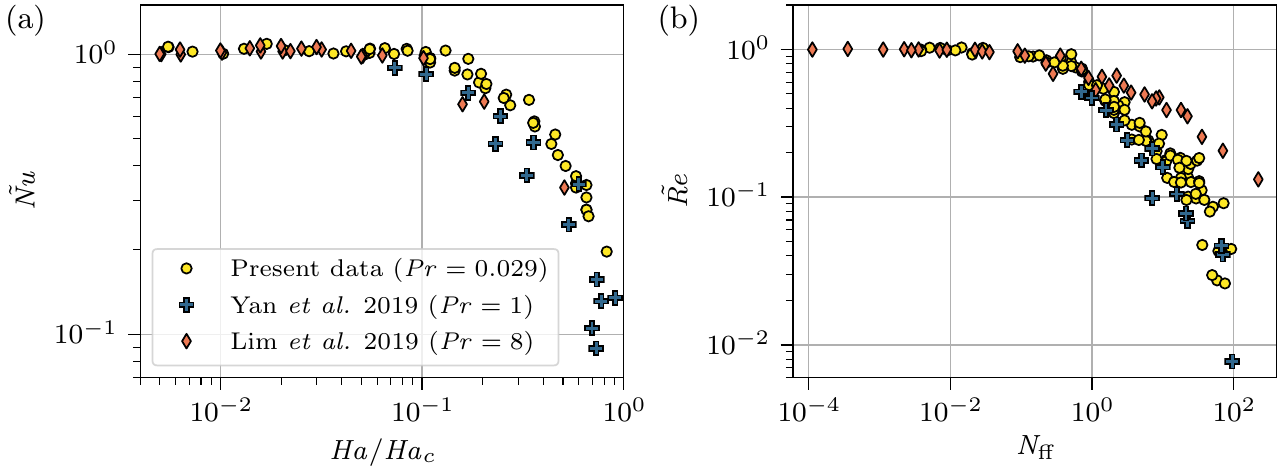}
\caption{%
  Comparison of the $\Pr$-dependence of the (a)~normalized Nusselt numbers over $\Ha/\Ha_\mathrm{c}$ and (b)~normalized Reynolds numbers over $\N_\mathrm{ff}$ with simulations by \citet{Lim2019} at $\Pr = 8$ and \citet{Yan2019} at $\Pr = 1$.
  The circles are the present data from figures~\ref{fig:Nu(Ha)} and~\ref{fig:Re(Nff)}.
  The numerical Reynolds numbers are based on the root mean square of the velocity field over the whole fluid volume.}
\label{fig:literature_Pr}
\end{figure}

DNS of magnetoconvection in vertical magnetic fields at moderate Prandtl numbers are reported by \citet{Yan2019} for $\Pr=1$ and \citet{Lim2019} for $\Pr=8$.
While \citet{Lim2019} use a cubic cell with solid walls, \citet{Yan2019} use boxes of different aspect ratios varying between $0.75:0.75:1$ and $28:28:1$ and with stress-free boundary conditions.
Figure~\ref{fig:literature_Pr} compares these numerical results with our low--Prandtl--number experimental data.
The data of $\tilde\Nu$ (figure~\ref{fig:literature_Pr}(a)) do not show a significant difference for the different Prandtl numbers, recalling the scatter of the experimental records at $\Pr = 0.025$ in figure~\ref{fig:literature_Nu}.
In contrast, the Reynolds numbers in figure~\ref{fig:literature_Pr}(b) reveal a dependence on~$\Pr$.
At first, the $\Pr = 0.029$ and $\Pr=1$ data have the same progression.
A fit of~\eqref{eq:Re_fit} to the $\Pr=1$ data results in parameters $\chi_2 = 1.19$ and $\gamma_2 = 0.76$.
However, the results for $\Pr = 8$ give a much slower decrease of $\tilde\Re$ with increasing $\N_\mathrm{ff}$ than the other data sets for $\Pr \le 1$ with fit parameters $\chi_2 = 0.44$ and $\gamma_2 = 0.51$.
Nevertheless, with the chosen normalization each data set consistently collapses onto an individual curve for their full Rayleigh number ranges which are $2\times10^4 < \Ra < 2\times 10^6$ for \citet{Yan2019} and $10^7 < \Ra < 10^{10}$ for \citet{Lim2019}.
Our initial ansatz~\eqref{eq:Re_argument} for the scaling parameter of $\tilde\Re$ included a factor of $1/\sqrt{\Pr}$, which was not relevant so far since the considered $\Pr$ were approximately constant.
If the $\tilde\Re$ data in figure~\ref{fig:literature_Pr}(b) were to be plotted over $\N_\mathrm{ff}/\sqrt{\Pr}$, the $\Pr = 1$ and 8 data would shift left of the $\Pr=0.029$ data by a factor of 0.2 and 0.06, respectively. 
In this case, however, the common start of the decrease of the Reynolds number would be lifted. 
Consequently, $N_\mathrm{ff}$ is chosen as the scaling parameter. 
This comparison of $\tilde\Re$ for different Prandtl numbers suggests that the parameters~$\chi_2$ and~$\gamma_2$ in the scaling~\eqref{eq:Re_fit} are a function of~$\Pr$, but not~$\Ha$.
The relation $\tilde\Re(\N_\mathrm{ff})$ seems to be mostly unchanged for $\Pr \le 1$, though the data displays a large scatter.
For $\Pr > 1$ the decrease of $\tilde\Re$ with $\N_\mathrm{ff}$ however becomes $\Pr$-dependent even if this observation is based on a single data set at $\Pr = 8$.

\section{Conclusion}
\label{sec:conclusion}

In this article, we present measurements of liquid metal Rayleigh-B\'enard convection in a cylindrical container of aspect ratio $\Gamma=1$ under the influence of an external vertical magnetic field of different field strengths.
The large-scale flow in the closed cylinder is investigated using temperature measurements at the side wall at half-height in combination with ten ultrasound Doppler velocimetry sensors.
It is shown that with increasing magnetic field strength, the initial large-scale circulation known from turbulent convection is first forced into a steady circulation mode by suppressing the regular oscillations of the torsion and sloshing mode.
A further increase of the magnetic field causes a breakdown of the one-roll structure into a cellular pattern of multiple convection rolls, which are identified using the spatially resolved velocity data.
Finally, the onset of the wall mode regime at Chandrasekhar's linear instability threshold is confirmed for a whole decade of $10^6 \le \Ra \le 10^7$:
The central flow is nearly fully suppressed and only in the vicinity of the side walls is a convective movement of the fluid still present.
This regime has received little consideration so far, but is now confirmed in our experiment, together with simulations~\cite{Liu2018} and pioneering theoretical studies by \citet{Houchens2002} and \citet{Busse2008}.

The momentum transport in the convective system is continually decreasing with increasing magnetic field throughout all flow regimes. 
The normalized Reynolds number is shown to be a function of the free-fall interaction parameter.
In contrast, the heat transport stays practically constant throughout most of the LSC regime due to suppression of turbulent oscillations and an increased coherence of the LSC flow.
An increase of the Nusselt number for an optimal Hartmann number as in large $\Pr = 8$ simulations~\citep{Lim2019} is not observed.
The decrease of the heat flux through the system sets in once the one-roll structure starts to break down into a cellular structure.
Our comparison with numerical data at higher Prandtl numbers show that the qualitative behaviour of the Hartmann-number-dependence of the global transport laws is similar to the present liquid metal flow case and that differences arise mainly for the momentum transport at high magnetic field strengths.

Open questions concern to our view mainly the flow dynamics in the cellular regime and the hysteresis effects of the pattern selection.
They would require a fine-resolved parameter survey in the region where the LSC and cellular regimes can coexist.
Such hysteresis-type switches between different large-scale flow states affect the transport capabilities of the liquid metal flow and can thus for example influence the cooling efficiencies in nuclear engineering devices.
Investigations of magnetoconvection in liquid sodium at even lower Prandtl numbers would thus also be desirable.

\begin{acknowledgments}
This work is supported by the Deutsche Forschungsgemeinschaft with Grants~No.\ GRK 1567, VO2331/1-1, and SCHU 1410/29-1.
\end{acknowledgments}

\bibliography{Magnetoconvection_bibliography}

\end{document}